# Ferroelectric Nanoparticles in Liquid Crystals: The Role of Ionic Transport at Small Concentrations of the Nanoparticles

Juliya M. Gudenko[1], Oleksandr S. Pylypchuk[1], Victor V. Vainberg[1], Igor A. Gvozdovskyy[1], Serhii E. Ivanchenko[2], Denis O. Stetsenko[1], Nicholas V. Morozovsky[1], Volodymyr N. Poroshin[1], Eugene A. Eliseev[2*], and Anna N. Morozovska[1†]

[1] Institute of Physics, National Academy of Sciences of Ukraine,
46, pr. Nauky, 03028 Kyiv, Ukraine

[2] Frantsevich Institute for Problems in Materials Science, National Academy of Sciences of Ukraine,
3, str. Omeliana Pritsaka, 03142 Kyiv, Ukraine

**Abstract**

We reveal the visible influence of the ultra-small concentrations (1 wt.% or less) of the $BaTiO_3$ nanoparticles (average size 24 nm) on the current-voltage characteristics and capacitance of the dielectric liquid crystal (LC) 5CB. The pure LC cell demonstrates higher current (and thus smaller resistance) than the LC cells filled with a very small concentration (0.5 – 1) wt.% of BTO nanoparticles. The same trend is observed for the charge-voltage characteristics: the capacitance loop is the widest for the pure LC cell and becomes noticeably thinner in the presence of (0.5 – 1) wt.% of $BaTiO_3$ nanoparticles. This seems counterintuitive, because 1 wt.% of ferroelectric nanoparticles very slightly modify the effective dielectric response and should not influence on the director distribution and elastic properties of the LC. We conclude that a possible physical reason of this observation is the influence of the ionic-electronic screening charges, which cover the ferroelectric nanoparticles and become polarized in the external field, on the ionic transport in the LC.

## 1. INTRODUCTION

Experimental implementation of ferroelectric nanoparticles (NPs) is abundant (see e.g., Refs. [1, 2, 3, 4, 5]), and the sizes of 5 – 50 nm are typical experimental values [6, 7, 8, 9]. There are several studies that use ferroelectric NPs fabricated in heptane and oleic acid producing core-shell nanoparticles [7, 10, 11, 12, 13].

---

[*] corresponding author, e-mail: anna.n.morozovska@gmail.com
[†] corresponding author, e-mail: eugene.a.eliseev@gmail.com

The ferroelectric NPs is very important for advanced optical, optoelectronic, and chemical applications [8 - 14, 15]. In the experimental papers [7, 10, 11, 15] particles suspended in a liquid dielectric medium were studied. These core-shell NPs provided a spontaneous polarization up to 130 μC/cm$^2$ [7, 11, 16]. Ferroelectric and paraelectric nanoparticles can influence significantly on the phase transitions and dynamics in nematic liquid crystals [12]. Despite the advances, many aspects still contain challenges for the preparation technology and unravelling the mystery [17] for the fundamental science [18, 19, 20]. In this context, small BaTiO$_3$ (BTO) NPs embedded in liquid or polymer matrices are of permanent scientific interest [21, 22, 23].

This work studies the influence of very small concentrations (1 wt.% or less) of the BTO NPs (average size 24 nm) on the current-voltage and charge-voltage characteristics, dielectric permittivity and losses of the dielectric liquid crystal (LC) 5CB.

## 2. EXPERIMENTAL RESULTS
### A. Samples preparation

To study the influence of NPs on the current-voltage characteristics the symmetrical LC cells filled by the pure nematic LC 5CB and liquid crystalline suspension based on BaTiO$_3$ NPs dissolved in the 5CB were used. At first, the glass substrates (microscope slides, made in Germany) were coated with thin ITO ((In$_2$O$_3$)$_{0.9}$ – (SnO$_2$)$_{0.1}$) layer (see **Fig. 1(a)**). The glass substrates with ITO layer were additionally covered by the n-menthyl-2-pyrrolidone solution of PI2555 (HD MicroSystems, USA) [24] in ratio 10:1 and spin-coated (6800 rpm over 10 s) to obtain a thin aligning film as shown in **Fig. 1(b)**. The substrates were dried at 80°C for 15 min, and further thermo-polymerization of polyimide molecules at 180 °C during 30 min was performed. To reach the strong azimuthal anchoring energy about 10$^{-5}$ J/m$^2$ [25] the PI2555 film was rubbed $N_{rubb}$ = 15 times (**Fig. 1(c)**).

LC cells were assembled of two rubbed substrates oriented in such a way that each of them have opposite rubbing direction with respect to another (**Fig. 1(d)**). The thickness of the LC cell, set by a spherical spacer of the 20 μm diameter, was 21 ± 0.5 μm. The thickness was measured by means of the interference method by recording the transmission spectrum of the empty LC cell using the Ocean Optics USB4000 spectrometer (Ocean Insight, USA, California). The LC cell was assembled of two substrates with dimension 15×10 mm$^2$. The liquid-crystalline suspensions were prepared using the nematic 4-pentyl-4'-cyanobiphenyl (5CB) and BaTiO$_3$ nanopowder (Nanotechcenter LLC, Ukraine) with a mean particle diameter of 24 nm. The nematic 5CB was synthesized in STC "Institute of Single Crystals" (Kharkiv, Ukraine) and purified before using. We studied the LC suspensions having 0.5 wt.% and 1 wt.% of BaTiO$_3$ NPs, added to the nematic 5CB. To obtain the uniform distribution of BaTiO$_3$ NPs in nematic 5CB the subsequent 30 min sonication of the suspension at the room temperature was carried out by means of the UZDN-2T ultrasonic disperser (Ukrrospribor, Sumy, Ukraine), at the frequency of 22 kHz and the output power of 75 W. The LC cell was filled using the capillary method at the temperature higher than the isotropic transition temperature of 5CB ($T_{Iso}$ = 308 K) to avoid the additional aligning of LC molecules owing to the flow (see **Figs. 1(e)** and **1(f)**).

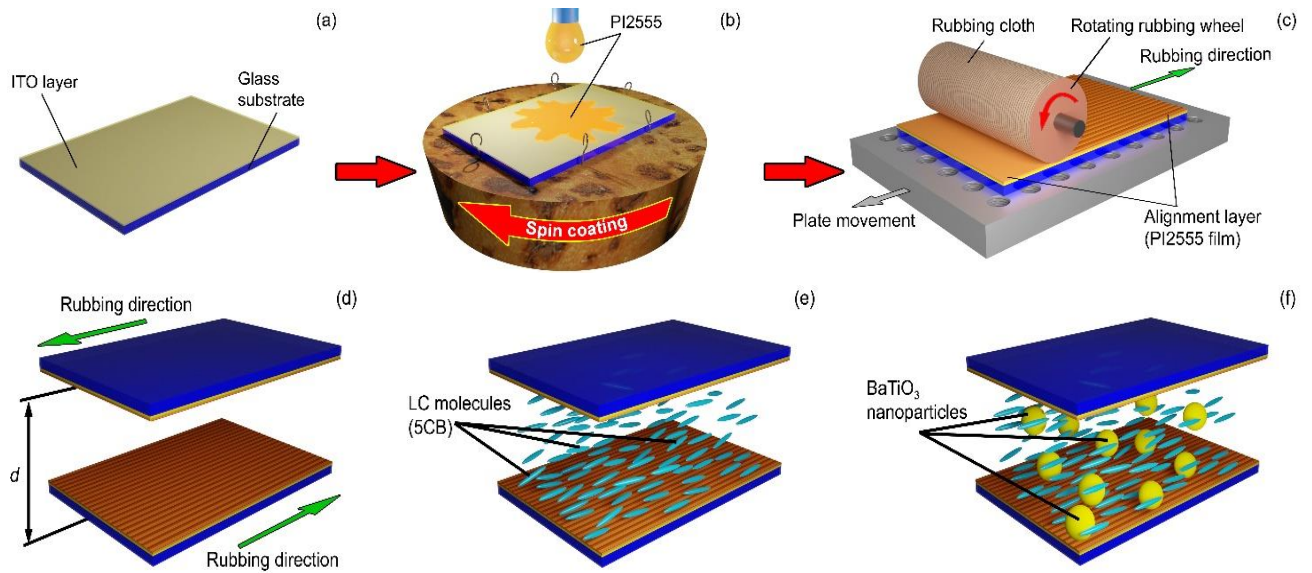

**Figure 1.** Schematic illustration of LC cells preparation stages **(a)-(d)**, the LC cell **(e)** and the LC cell with BTO NPs **(f)**.

## B. Electrophysical measurements

The electric scheme for measurements of the steady state electric transport characteristics and transient processes in the studied samples is shown in **Fig. 2**. The electric voltage on the sample with a series-connected resistor was applied from the software-controlled voltage supply GW Instek PSP-603. To eliminate possible transient delays at measuring steady state current-voltage characteristics the voltage magnitude was slowly varied with the rate of 20 mV per 5s. The voltage drops on the sample with the load resistor were measured the by Keithley-2000 digital multimeters and recorded by the control computer. The Aktakom ACK-3106 digital oscilloscope in recorder mode was also used to record fast transients instead of the multimeters. To measure the volt-farad characteristics and capacitance, the RLC meters UNI-T UT612 (100, 120, $10^3$, $10^4$ and $10^5$ Hz) and E7-12 ($10^6$ Hz), were used in conjunction with an external software-controlled dc voltage supply GW Instek PSP-603. The electrical contacts to the LC + BaTiO$_3$ cell were made by soldering indium onto the glass surface covered with an ITO layer.

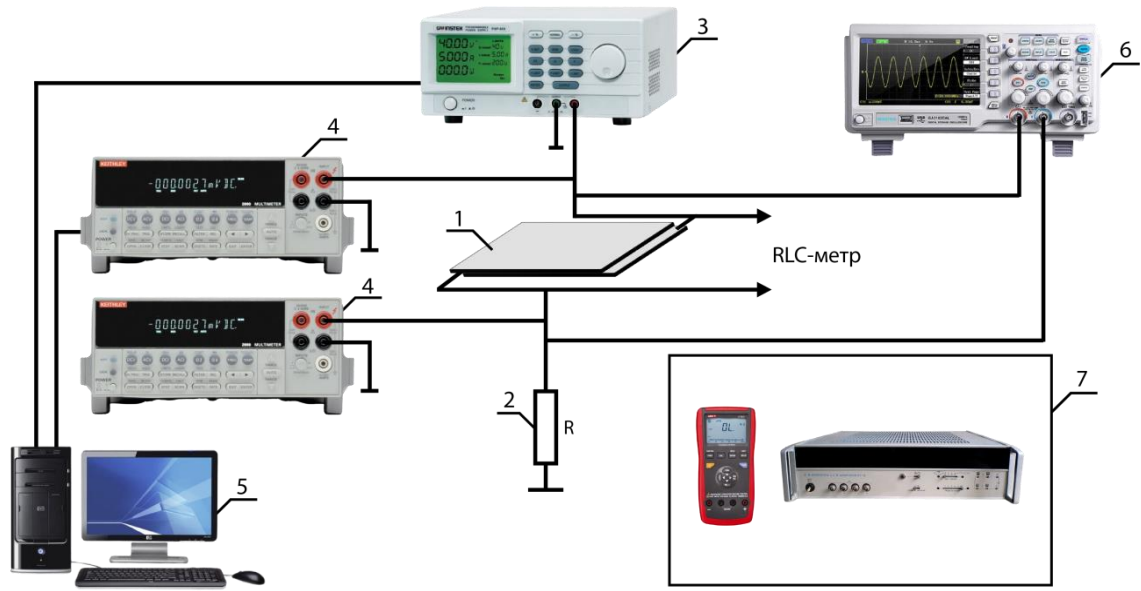

**Figure 2.** Scheme of measurements of the electrical characteristics: 1 – the LC+BTO cell sample, 2 – the load resistor for measuring the current through the sample; 3 – voltage supply GW Instek PSP-603; 4 - Keithley 2000 digital multimeters, 5 - control and recording computer, 6 - digital oscilloscope Aktakom ACK-3106 for measuring fast transients, 7 - RLC meters: UNI-T UT612 and E7-12 for measuring the capacity of the sample.

The inset in **Fig. 3** depicts the TEM image to show the actual sizes and shapes of the BTO NPs and presenting some information about the spread of these parameters. As it is seen, the shape can be regarded spherical in average. The average size (24 nm) of the prepared NPs, as well as the spread of their sizes (from 17 nm to 47 nm) were determined by a conventional way [21]. The phase composition was determined by the Rietveld refinement as 7 wt.% of orthorhombic $BaCO_3$ and 93 wt.% of tetragonal $BaTiO_3$ [21].

The current-voltage characteristics of the pure LC cells and the LC cells with 0.5 wt.% and 1.0 wt.% of BTO NPs measured in the dc regime are shown in **Fig. 3**. There appeared that the empty LC cell demonstrates higher current at the same voltage bias (and thus smaller resistance) than the LC cells filled with a very small concentration (0.5 wt.%) of BTO NPs (see **Fig. 3**). The increase of the NPs concentration to 1 wt.% leads to further small decrease of the current especially at higher voltages. All dc current-voltage characteristics in **Fig. 3** manifest similar loop-like behavior, which is analyzed in more detail for one of them in **Fig. 4**.

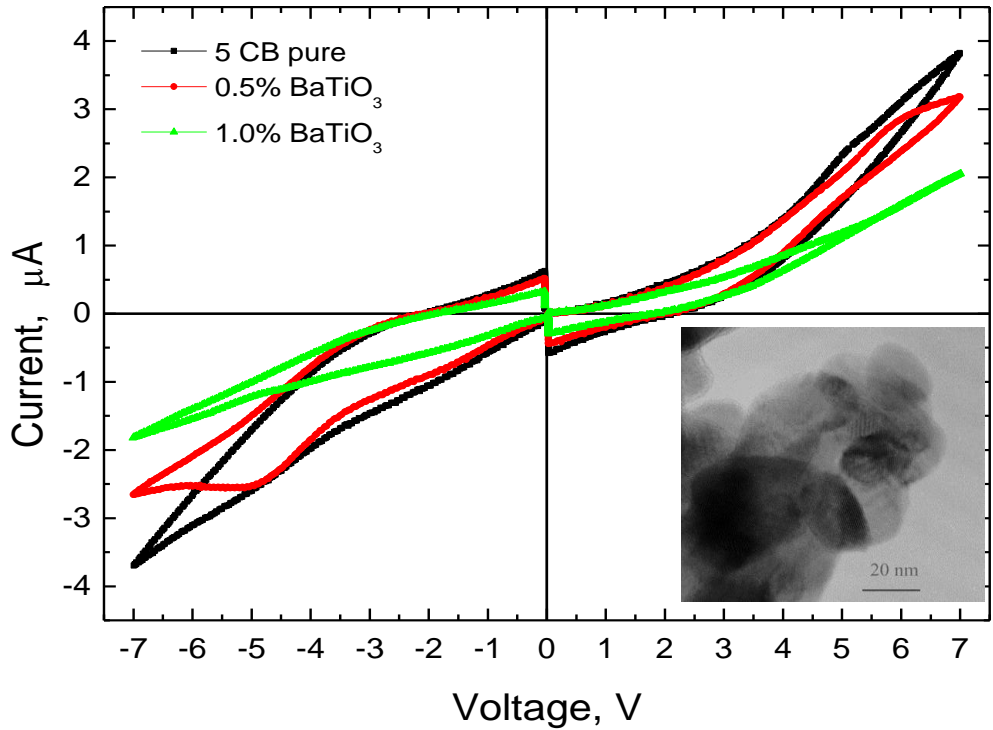

**Figure 3.** Current-voltage characteristics of 3 LC cells with different content of BTO NPs measured at the room temperature (300 K). Black loop: the cell with a pure LC 5 CB (thickness 20 μm, sample 1), red loop: the LC cell with 0.5 wt.% BTO NPs (thickness 20.7 μm, sample 2), green loop: the LC cell with 1.0 wt.% BTO NPs (thickness 20.6 μm, sample 3).

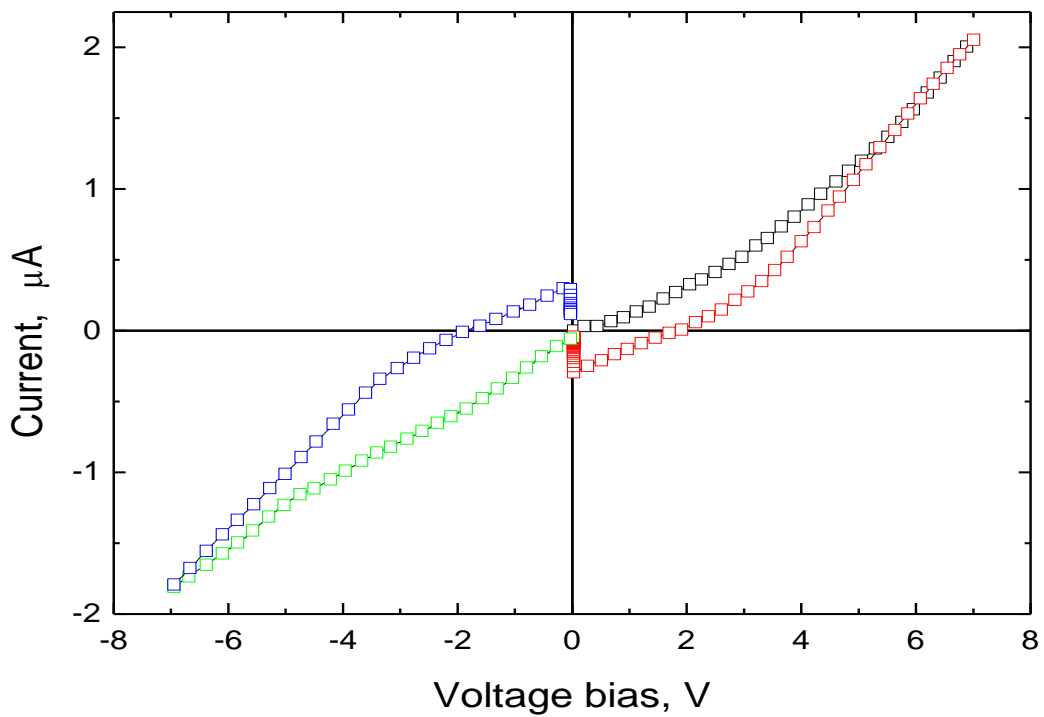

**Figure 4.** Current-voltage characteristic of the LC cell with 1 wt.% of BTO NPs at 300 K. The black and green curves correspond to increasing voltage bias, both in the forward and the backward

polarity, the red and blue curves correspond to decreasing bias, respectively.

It is seen that the branch at the increasing voltage bias has a super linear growth at both polarities, while beginning from the zero current. The backward branches at decreasing voltage bias manifest quicker decrease with entering at some voltage to the field of the opposite current direction, remaining the opposite sign voltage across the sample, about 3 mV, at the zero voltage from the supply unit. This evidently says of remaining opposite sign polarization of the sample which corresponds to the electric field about 1.5 V/cm. After decreasing voltage bias down to 0 the sample remained unbiased for about 10 minutes to discharge in its circle with the load resistor of 10 kOhm. After decreasing the current down to 0 the loop was recorded with the opposite sign bias and in general it almost repeats the initial loop shape.

The same trend is observed for the charge-voltage characteristics: the capacitance loop is the widest for the empty LC cell, and it becomes noticeably thinner in the presence of (0.5 – 1) wt.% of BTO NPs (see **Fig. 5**). This seems counterintuitive, because 1 wt.% of the NPs very slightly modifies the mixture dielectric constant. A possible physical reason of this observation is the influence of the ionic-electronic screening charges and dielectrophoretic forces (which are very weak but at the same time long-range and omnipresent) on the ionic transport in the LC.

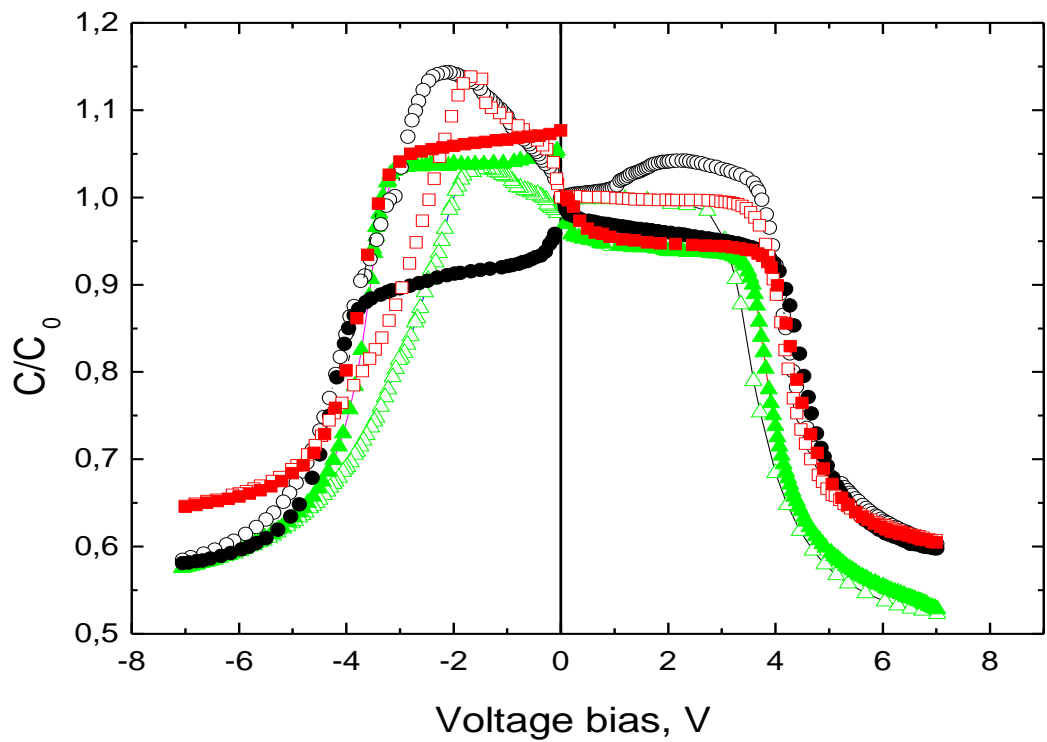

**Figure 5.** The dependence of the ratio $C/C_0$ ($C$ is the capacitance, $C_0$ is a zero-bias capacitance) vs. the bias voltage of the LC cell with BTO NPs measured at the room temperature (300 K). Black circles correspond to the cell with a pure LC 5CB (thickness 20 μm, sample 1), red squares

correspond to the LC cell with 0.5 wt.% BTO NPs (thickness 20.7 μm, sample 2), green triangles correspond to the LC cell with 1.0 wt.% BTO NPs (thickness 20.6 μm, sample 3). The empty symbols correspond to increasing bias, filled symbols correspond to decreasing bias.

In contrast to the current-voltage characteristics, the capacitance vs the voltage bias curves (except the feature mentioned above) have an antisymmetric shape in regard of the polarity change. The capacitance keeps its value close to that at the zero bias in the bias range -4 to +4 V. At higher bias, which correspond to field more than 2 kV/cm, the capacitance sharply decreases in two times and then tends to saturation. Such behavior is typical, for example, to the situation when some fraction of space charge contributing to slow polarization is eliminated by a strong electric field.

The frequency dependences of the dielectric permittivity and losses of the pure LC cell and LC cells with 0.5 wt.% and 1 wt.% BTO NPs are shown **Fig. 6(a)** and **Fig. 6(b)**, respectively. As it is seen in **Fig. 6(a),** the dielectric permittivity of all samples monotonously decreases with the increasing frequency in the whole range from 100 Hz to 1 MHz. In the range of frequencies ($10^2 – 10^5$) Hz the decrease is slow and at higher frequencies one observes a quick decrease. The capacitance of the pure LC cell is the smallest and the capacitance of the LC cell with 0.5 wt.% and 1 wt.% of the BTO NPs in fact coincides with each other in the entire frequence range. The weak increase of the low-frequency permittivity with addition of the BTO NPs concentration is in a qualitative agreement with the Maxwell-Garnett effective media approximation. A fast decrease in the permittivity is observed for the frequences above 0.1 MHz; and all permittivity curves merge in the frequency range. The significant (by 10 times) decrease in the capacitance with the frequency increase over 4 orders of magnitude is inherent to the ions motion in the LC and hardly possible for the wide-gap ferroelectric materials.

As it is seen in **Fig. 6(b),** the losses of the pure LC cell and LC cells with 0.5 wt.% and 1 wt.% BTO NPs vary non-monotonously, at first decreasing with the frequency increase from $10^2$ Hz to $10^4$ Hz and then increasing with frequency. At $10^4$ Hz the losses of all cells reach the absolute minima. In general, the dependences of the losses angle tangent on the frequency are very close to each other, which evidences on the prevailing contribution of the LC matrix. While the decreasing with frequency part may be related with a decreasing ratio of real and imaginary components of the complex impedance, the increasing part at higher frequencies may be related to the retarding effects in the screened particles.

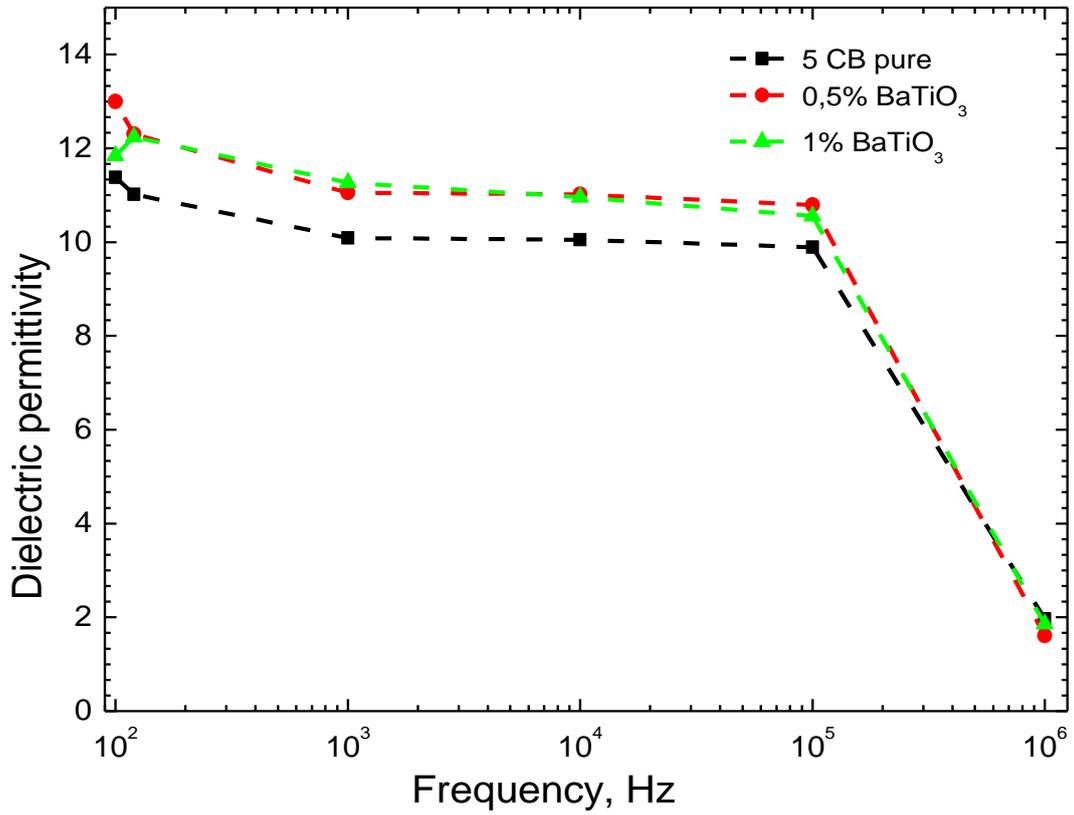

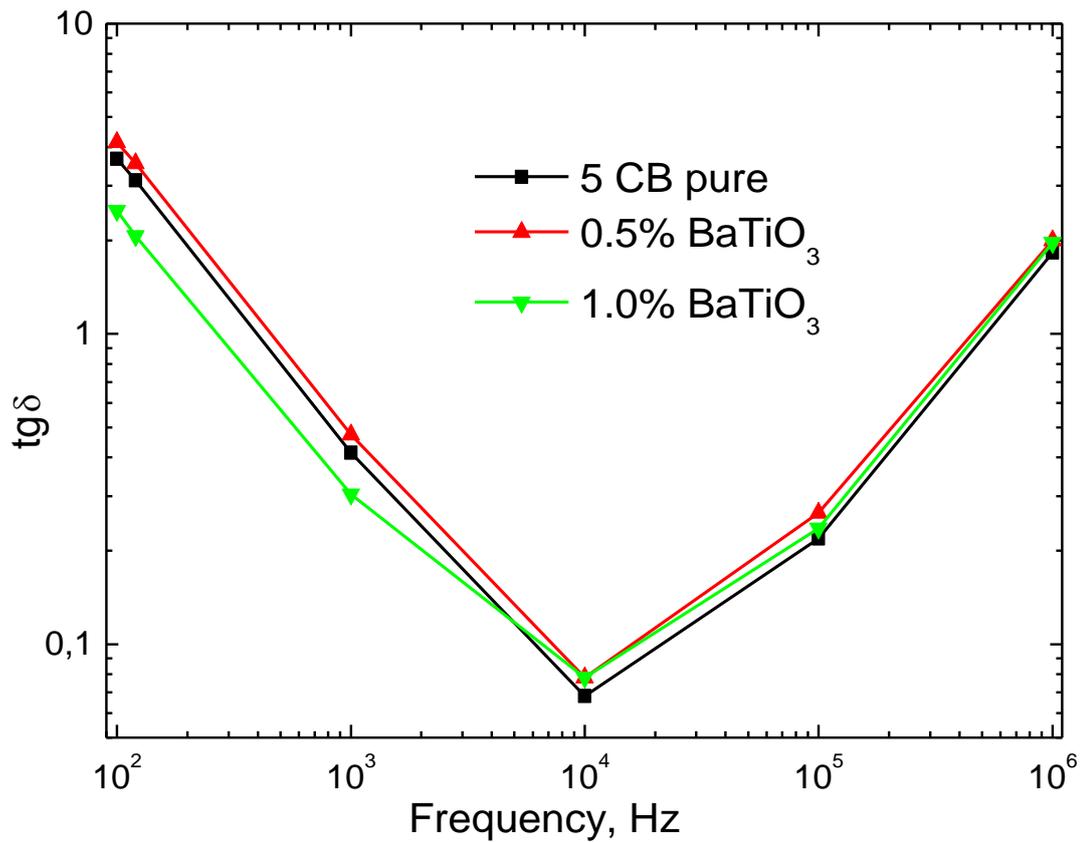

**Figure 6.** Effective dielectric permittivity (**a**) and losses angle tangent tgδ (**b**) of the LC cells with BTO NPs measured at room temperature (300 K). Samples description is the same as in **Fig. 3**.

We also measured the current-voltage characteristics in the frequency range 10 Hz – 1 kHz using a triangular-shaped applied voltage. The current-voltage characteristics correspond to the linear RC circuit at 1 kHz, to the RC circuit with a very small nonlinearity at 100 Hz, and to the RC circuit with more pronounced nonlinearity (Rayleigh bi-angle like loop shape) at 10 Hz. The multiplicity of resistance increase is (1,3 - 2) times and is maximal for 5CB with 1 wt.% of BTO NPs, which resistance is maximal. The multiplicity of the capacitance increase is many more times than the same of the resistance and the capacity increase is maximal (~ 10 times) under transition from 100 Hz to 10 Hz. This result well agrees with the conclusion that the slow dynamics of the ionic-electronic charge, which screens the ferroelectric NPs and/or accumulated near the electrodes, determines the nonlinear features of the current-voltage curves of the LC cell with or without ferroelectric NPs.

## 3. THEORETICAL MODEL

It is also well known that the small NPs (size less than 10 – 50 nm) with concentrations less than 1 wt.% cannot change the director distribution and/or elastic properties of the LC in a noticeable way. Hence, the small concentrations of the NPs cannot influence in a direct way on the current-voltage and charge-voltage characteristics of the LC cells. However, the above-presented experimental results reveal the visible influence of the ultra-small concentrations of the BTO NPs (1 wt.% or less) on the electrophysical properties and effective capacitance of the dielectric LC 5CB. A possible physical reason of this observation is the indirect influence of the screening charges, which cover the ferroelectric NPs and become polarized in the external field, on the slow ionic transport in the LC. Below we use the effective media approach to calculate the effective capacitance of the LC 5CB and colloid LC 5CB with BTO NPs.

### A. Effective media approximation

There are many effective media approaches (shortly "EMA") [26], among which the most known are the Landau approximation of linear mixture [27], Maxwell-Garnett [28] and Bruggeman [29] approximations for spherical inclusions, and Lichtenecker-Rother approximation of logarithmic mixture [30]. Most of these approximations are applicable for quasi-spherical randomly distributed dielectric (or semiconducting) particles in the insulating environment. These EMA models lead to the evident expression for the effective permittivity $\varepsilon_{eff}^*$. However their applicability ranges are very sensitive to the cross-interaction effects of the polarized NPs, and therefore most of them can be invalid for dense composites and/or colloids, where the volume fraction of ferroelectric particles is more than (20 – 30) %.

The effective media approximation, also abbreviated as "EMA", was proposed by Carr et al. [31]. EMA was successfully used by Petzelt et al. [32] and Rychetský et. al. [33]. It gives the quadratic equation for the effective permittivity of the binary mixture:

$$(1-\mu)\frac{\varepsilon^*_{eff}-\varepsilon^*_b}{(1-n_a)\varepsilon^*_{eff}+n_a\varepsilon^*_b} + \mu\frac{\varepsilon^*_{eff}-\varepsilon^*_a}{(1-n_a)\varepsilon^*_{eff}+n_a\varepsilon^*_a} = 0. \quad (1)$$

Here $\varepsilon^*_a$, $\varepsilon^*_b$ are relative complex permittivity of the components "a" and "b" respectively, $\mu$ and $1-\mu$ are relative volume fractions of the components "a" and "b" respectively, and $n_a$ is the depolarization field factor for the inclusion of the type "a".

Using the EMA for $\mu \ll 1$ (and hence $\varepsilon^*_{eff} \approx \varepsilon^*_b$) in Ref. [34] we derived the linearized equation for the effective permittivity $\varepsilon_{eff}$:

$$\varepsilon^*_{eff} = \varepsilon^*_b\left[1 + \frac{\mu(\varepsilon^*_a-\varepsilon^*_b)}{n_a\varepsilon^*_a+(1-n_a)\varepsilon^*_b-n_a\mu(\varepsilon^*_a-\varepsilon^*_b)}\right]. \quad (2a)$$

Here we supposed that the external electric field is pointed along one of the ellipsoidal particles principal axes, for which the depolarization factor is $n_a$. It is seen that Eq.(2a) contains the term $-n_a\mu(\varepsilon^*_a - \varepsilon^*_b)$ in the denominator, which makes Eq.(2a) more consistent with the solution of the EMA equation (1). Similar expression (but without the term) was derived by Hudak et al. [35].

It is seen that Eq.(2a) is the first two terms of the expansion of the effective permittivity (1) with respect to $\mu$. To study adequately an anisotropic liquid, as a surrounding medium, such as liquid crystals, we need to consider the effects of the $\varepsilon_e$ anisotropy, which can be rather strong (see e.g., Refs [36, 37]). Next, we assume that the NPs have the complex dielectric permittivity $\varepsilon^*_a = \varepsilon_a - i\frac{\sigma_a}{\omega}$ and the LC has the complex permittivity:

$$\varepsilon^*_b = \varepsilon_\infty + \frac{\varepsilon_b-\varepsilon_\infty-i\frac{\sigma_b}{\omega}}{1+i\,\tau\,\omega}. \quad (2b)$$

The LC 5CB parameters can be found in Ref.[38].

Frequency dependences of the real part of the dielectric permittivity $\text{Re}(\varepsilon^*_{eff})$, losses angle tangent $tg\delta$, and losses $\text{Im}(\varepsilon^*_{eff})$, are shown in **Fig. 7(a)-7(c)**. The dependences are calculated using Eqs.(2) for the parameters $n_a = 0.33$, $\varepsilon_a = 500$, $\varepsilon_\infty = 1$ and $\varepsilon_b = 10$. The values of the parameters $\sigma_a = 1$ s$^{-1}$, $\tau = 1.91 \cdot 10^{-6}$ s and $\sigma_b = 5.1 \cdot 10^3$ s$^{-1}$ were determined from the best agreement with experimental dependences shown in **Fig. 6.**

The calculated frequency dependences of the real part of the dielectric permittivity are in the quantitative agreement with experimentally measured (compare **Fig. 6(a)** with **Fig. 7(a)**). The effective permittivity of the LC cell with 1.0 % of BTO NPs is slightly higher than the permittivity of the LC cell with 0.5 % of BTO NPs, and the latter is slightly higher than the permittivity of a pure LC 5 CB (compare the green, red and black curve in **Fig. 7(a)**).

The calculated frequency dependences of the tgδ are in the semi-quantitative agreement with experimentally measured (compare **Fig. 6(b)** with **Fig. 7(b)**). The tgδ of the LC cell with 1.0 % of BTO NPs, the LC cell with 0.5 % of BTO NPs, and of the pure LC 5CB coincides (compare the green, red and black curve **Fig. 7(b)**). The tgδ reaches the minimum near 20 kHz. The parameters $\sigma_b \gg \sigma_a$ and $\tau$ determine the frequency position of the minimum. The minimum is very weakly sensitive to the parameters $\sigma_a$ (until $\sigma_a \ll \sigma_b$), $\varepsilon_a$ and $\varepsilon_b$.

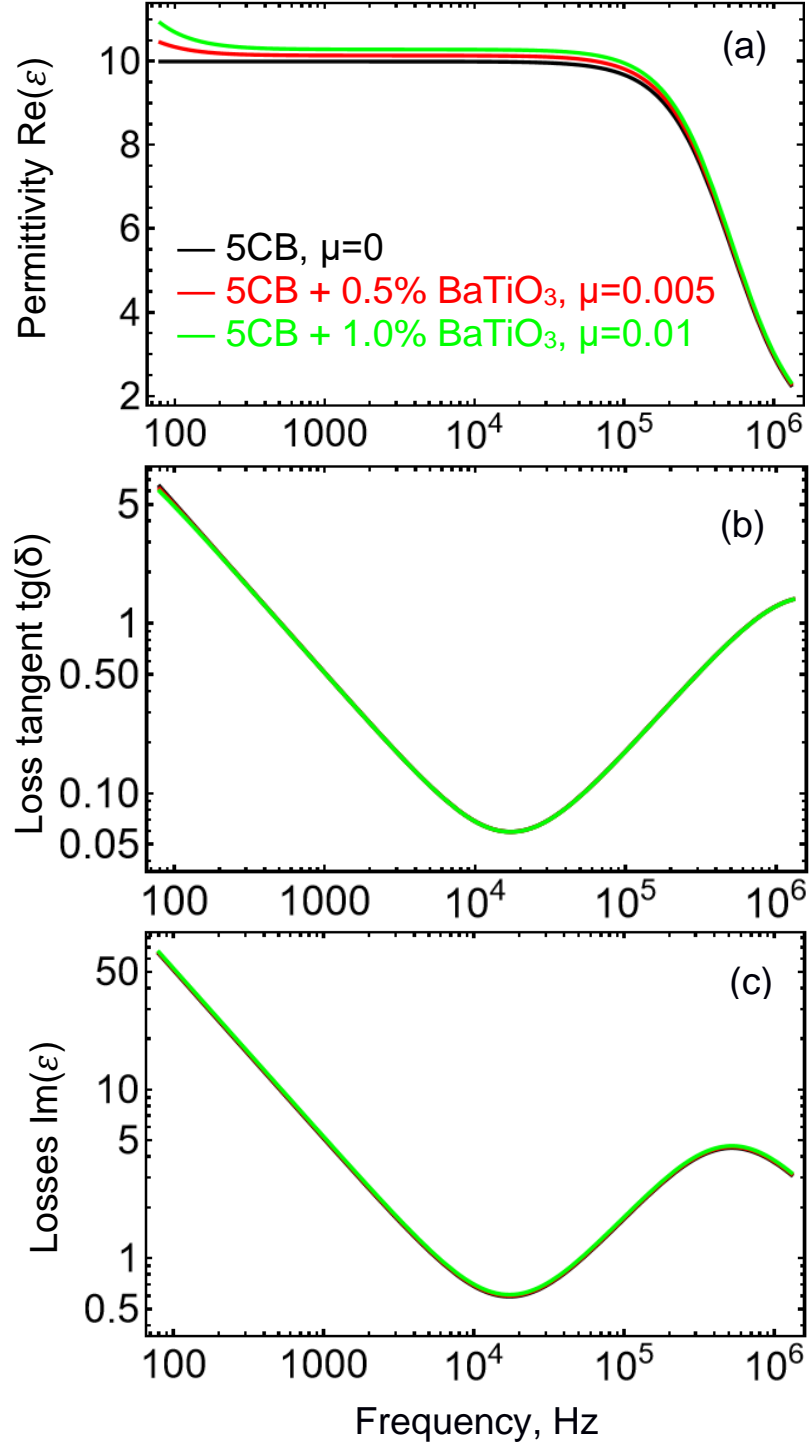

**Figure 7.** Frequency dependences of the real part of the dielectric permittivity **(a)**, losses angle tangent tgδ **(b)**, and imaginary part of dielectric permittivity **(c)**, calculated from Eqs.(2) for $n_a = 0.33$, $\varepsilon_a = 500$, $\sigma_a = 1$ s$^{-1}$, $\varepsilon_\infty = 1$, $\varepsilon_b = 10$, $\tau = 1.91 \cdot 10^{-6}$ s and $\sigma_b = 5.1 \cdot 10^3$ s$^{-1}$. Here the relative volume fraction $\mu = 0$ for the pure LC 5 CB, $\mu = 0.005$ for the LC cell with 0.5 % of BTO NPs, and $\mu = 0.01$ for the LC cell with 1.0 % of BTO NPs.

The calculated frequency dependences of the losses are the same in the LC cell with 1.0 % of BTO NPs, the LC cell with 0.5 % of BTO NPs, and of the pure LC 5CB coincides (compare the green, red and black curve in **Fig. 7(c)**).

## 4. Conclusions

We reveal the visible influence of the very small concentrations (1 wt.% or less) of the $BaTiO_3$ NPs (average size 24 nm) on the current-voltage characteristics and capacitance of the LC 5CB. The pure LC cell demonstrates higher current (and thus smaller resistance) than the LC cells filled with the (0.5 – 1) wt.% of BTO NPs. The same trend is observed for the charge-voltage characteristics: the capacitance loop is the widest for the pure LC cell and becomes noticeably thinner in the presence of (0.5 – 1) wt.% of BTO NPs. This seems unusual, because 1 wt.% of the NPs very slightly modify the effective dielectric constant and practically do not influence on the director distribution and elastic properties of the LC.

A possible physical reason of this is the influence of the screening charges, which cover the ferroelectric NPs and become polarized in the external field, on the slow ionic transport in the LC. To quantify the experimental results, we calculate the effective dielectric permittivity and losses for an ensemble of the NPs in liquid dielectric media in the effective media approximation. Analytical calculations are in the semi-quantitative agreement with experimental results. The obtained results can be used to tailor metamaterials for multiple practical applications.

**Acknowledgements.** The work of J.M.G, O.S.P., E.A.E. and A.N.M. are funded by the National Research Foundation of Ukraine (project "Manyfold-degenerated metastable states of spontaneous polarization in nanoferroics: theory, experiment and perspectives for digital nanoelectronics", grant N 2023.03/0132 and "Silicon-compatible ferroelectric nanocomposites for electronics and sensors", grant N 2023.03/0127). The work of V.V.V., N.V.M. and V.N.P. are funded by the Target Program of the National Academy of Sciences of Ukraine, Project No. 4.8/23-p. The materials preparation characterization (S.E.I.) is sponsored by the NATO Science for Peace and Security Programme under grant SPS G5980 "FRAPCOM".